\documentclass[%
 reprint,
superscriptaddress,
showpacs,preprintnumbers,
nofootinbib,
 amsmath,amssymb,
 aps,
pra,
]{revtex4-1}
\usepackage{verbatim}
\usepackage{textcomp}
\usepackage{color}
\usepackage{graphics}
\usepackage{filecontents}
\usepackage{graphicx}
\usepackage{dcolumn}
\usepackage{bm}
\usepackage{hyperref}
\usepackage{natbib}

\begin{document}
\bibliographystyle{apsrev4-1}

\title{Effects of linear and quadratic dispersive couplings on optical squeezing in an optomechanical system }

 
 \author{Satya Sainadh U}
 \email{sainadh.undurti@griffithuni.edu.au}
 \affiliation{ Center for Quantum Dynamics, Griffith Universiy, Brisbane, QLD 4111, Australia.}
 \affiliation{ Raman Research Institute, Bangalore, India-560080.}
\author{M. Anil Kumar}%
 \email{anilk@rri.res.in}
\affiliation{ Raman Research Institute, Bangalore, India-560080.}%

\date{\today}
\begin{abstract}
A conventional optomechanical system is composed of   a mechanical mode and an optical mode interacting through a linear optomechanical coupling (LOC). We study how the presence of quadratic optomechanical coupling (QOC) in the conventional optomechanical system affects the system's stability and optical quadrature squeezing. We work in the resolved side-band limit with a high quality factor mechanical oscillator. In contrast to the conventional optomechanical systems, we find that strong squeezing of the cavity field can be achieved in presence of QOC along with LOC at lower pump powers and at higher bath temperatures. Using detailed numerical calculations we also find that there exists an optimal QOC where one can achieve maximum squeezing.
\end{abstract}
\maketitle
\section{Introduction}
\begin{figure}[h!] \centering
 \includegraphics[scale=0.55]{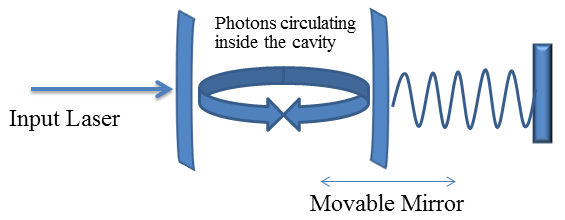}
 \caption{\label{fig1}{\color {black}(Color online)} Schematic of a cavity optomechanical system}
 \end{figure}

Recent advances in optomechanics have led to a deeper understanding of quantum features at macroscopic scales. A prototypical cavity optomechanical system, is represented by a single mode Fabry-Perot cavity with one movable end mirror as shown in Fig. \ref{fig1}. The mean position of the movable mirror is controlled by radiation pressure force of the light intensity circulating inside the cavity. {\color {black}The circulating intensity mediates an interaction between the cavity and mechanical degrees of freedom.} Most of the common studies as given in \citep{aspelmeyer} and references there in, considered linear optomechanical coupling (LOC) in which the cavity mode couples to the position of the mechanical mirror linearly. {\color {black}These LOC interactions are mostly used for quantum ground state cooling \cite{pinard,wilsonrae,*marquardt2007quantum,chan2011laser,*teufelnature} of the mechanical mirror, entanglement between light and the mirror \cite{vitali2007optomechanical,*palomaki}, electromagnetically induced transparency (EIT) \citep{agarwal,*EIT-slowlight}, optomechanical induced transparency (OMIT) \citep{kippenbergomit} and studies concerning normal-mode splitting \citep{dobrindt2008parametric,groblacher2009observation}.} 

However interactions with quadratic optomechanical coupling (QOC) have also been considered where the optical cavity mode is coupled to the square of the position of the mechanical oscillator. {\color {black}In such systems the movable mirror is replaced by a membrane \citep{j-d-thompson} or  ultra cold atoms \citep{purdy}}. The QOC interaction in the membrane in the middle of the cavity system has been used to observe quantization in mechanical energy \citep{j-d-thompson}, traditional two-phonon laser cooling \citep{nunnenkamp-squeezing}, tunable slow light \citep{tunablezhan}, photon blockade \citep{liao2013photon}, optomechanics at a single photon level \citep{franconori2014}, macroscopic tunnelling and quantum Zeno effect in optomechanical double well potential \citep{meystre-potential} etc. 

Though optomechanical systems offer a platform for wide variety of experimental and theoretical investigations, they are mostly studied with either LOC or QOC interactions alone. {\color {black}Recently, optomechanical systems with both LOC and QOC together were theoretically studied where the static response of the mechanical mirror was extremely sign sensitive to the QOC interaction \citep{quadratic}, squeezing and cooling of dielectric nano- or micro-spheres \citep{xuereb} inside an optomechanical system and harmonic generation of self-sustained oscillations in a cavity field \citep{zhang2014self}. Such a system  has also been experimentally demonstrated  where the mechanical resonator is prepared and detected near its ground state motion \citep{mechanicalground}.}
 
Along with the above mentioned studies, optomechanical systems also exhibit optical quadrature squeezing. Under radiation pressure force an optomechanical system in its steady state mimics a kerr medium \citep{fabre1994quantum} and the fluctuations in the system can give rise to optical quadrature squeezing, termed as ponderomotive squeezing
\citep{mancini,fabre1994quantum,marino}. It has wide applications in quantum limited displacement sensing \citep{braginskybook,Hoff}, large scale gravitational wave observatories \citep{eberle,ligo} and even in biological  measurements \citep{biological}. 

Using LOC interaction, squeezed light from cavity output has been studied and experimentally realized in an unresolved side-band regime \textit{i.e.} $\omega_m<\kappa$ around the resonant frequency of the mechanical oscillator. The experiment carried by D. W. C. Brooks \textit{et.al.} \citep{brooks} demonstrated that the back action of the motion of an ultracold atomic gas  on the cavity light field produces ponderomotive squeezing whereas, Safavi-Naeini \textit{et.al.} \citep{silisonmicroresonator} using silicon  micromechanical cavity resonator, observed fluctuation spectrum at a level 4.5$\pm $0.2\% below the shot-noise limit in highly excited thermal state ($10^4$ phonons). Soon after that Purdy et al. \citep{regalPRX} placed a semi-transparent membrane (made of silicon nitride) inside an optical cavity and observed a squeezing of  32\% (1.7dB) when membrane was cooled to about 1 mK. While in all the above  experiments squeezing was observed in MHz frequency range, A. Pontin \textit{et.al.} \citep{noisecancel} detected squeezed light in the audio band (kHz) at a bath temperature of 4 K, by almost completely cancelling the frequency noise. 
 
On the other hand S. M. Girvin and their co workers theoretically analysed and proposed quadrature squeezing \citep{nunnenkamp-squeezing}  with QOC interaction alone in the resolved side-band regime \textit{i.e.} $\omega_m>\kappa$. They showed that the system maps onto a degenerate parametric oscillator, where the cavity mode is coupled parametrically to the square of the position of a mechanical oscillator. This system facilitates optical squeezing around the frequencies $\omega=0,\pm2\omega_m$ at a very low phonon number. 

 Given the importance of optical quadrature squeezing in the field of metrology, it is very important for one to answer how much LOC and QOC together can affect squeezing. Therefore in this work, we aim to investigate the achievable level of quadrature squeezing of the cavity output light in presence of both LOC and QOC interactions together in a resolved side-band limit and strong photon-phonon coupling regime. Our scheme differs from the above mentioned schemes in the way that the addition of QOC modifies  the single photon optomechanical coupling to an intensity dependent coupling that aids in an increase in the magnitude of quantum fluctuations of the input field over thermal noise, which can enhance optical squeezing. Also the radiation pressure driving the mechanical oscillator modifies its resonance frequency considerably such that the squeezing can be observed around this modified frequency. This considerable modification of the mechanical oscillator's frequency can be seen as a signature of the QOC's presence. Such modification of frequency and enhanced optomechaical coupling was studied in \citep{crosskerr}. Unlike various proposals of hybrid optomechanical systems like weak Kerr non-linearity with degenerate OPA \citep{shahidani}, quantum well \citep{eyob}, and electromechanical systems with super conducting qubits \citep{electromechanical}, our scheme can be useful to observe strong squeezing that can be implemented in a simple cavity opto-mechanical system with no requirement of any extra degrees of freedom. 

The paper is organized as follows: Sec. II presents the theory describing the system with Hamiltonian, equations of motion, and the steady state values. Further we explain the dynamics of the system giving us the  stability criteria and derive the expressions for spectra of cavity output field and optical quadrature squeezing. Sec III deals with  numerical results and discussion regarding the effect of QOC on optical quadrature squeezing in presence of thermal fluctuations. Finally conclusions are given in Sec. IV.

\section{Theory}
In this section we introduce our system Hamiltonian and formulate the corresponding Heisenberg-Langevin equations of motion.
 We consider the optomechanical
 system of Fig. \ref{fig1}, with a single cavity mode frequency $\omega_c$ 
and cavity decay rate $\kappa$  coupled  dispersively to a single mechanical mode of
 frequency $\omega_m$. The interactions are both linear $(g_{_{1}})$ and quadratic $(g_{_{2}})$ in mechanical oscillator's displacement. 
The system is driven by a strong pump field of frequency $\omega_p$ . The complete Hamiltonian 
 of the system in the laser frame is given by
 \small
 \begin{equation}
 H=\hbar\Delta a^\dagger a+\frac{\hbar \omega_m }{2}(x^2+p^2)+\hbar g_{_{1}} a^\dagger a x+\hbar g_{_{2}} a^\dagger a x^2+i\hbar\varepsilon(a^\dagger-a). \label{a1}
 \end{equation}\normalsize
{\color {black}Here $\Delta=\omega_c-\omega_p$ is the cavity detuning and 
$a (a^\dagger)$ is the annihilation (creation) operator of the cavity mode such that $[a,a^\dagger ]  = 1$}. $x$ and $p$ refer
to the dimensionless position and momentum operator for the mechanical oscillator with the commutation relation as $[x,p]=i$.
In Eqn. (\ref{a1}), the first two terms represent the free energy of cavity and mechanical oscillator respectively,
 with the third and fourth terms representing the optomechanical interaction terms which arise due to the fact that the 
mechanical oscillator couples to the cavity field via its displacement both linearly and quadratically.
 $g_{_{1}}$ and $g_{_{2}}$ are the coupling constants associated with these interaction terms.
They are defined as LOC, $g_{_{1}}=\frac{\partial \omega_c}{\partial x} x_{zpf}$ and QOC, $g_{_{2}}=\frac{\partial^2 \omega_c}{\partial x^2} \frac{x_{zpf}^2}{2}$ \citep{aspelmeyer}. 
Here $x_{zpf}$ is the zero-point fluctuations of the mechanical oscillator's displacement given by $\sqrt{\frac{\hbar}{m \omega_m}}$ where $m$ 
is the effective mass of the oscillator. The last term describes the interaction of the cavity mode with the pump field amplitude 
$(\varepsilon=\sqrt{\frac{2\kappa \mathcal{P}}{\hbar \omega_p}})$, with $\mathcal{P}$ being the input power of the pump field. 
In order to fully describe the dynamics of the system
 it is essential to include fluctuation and dissipation processes
affecting the optical and mechanical modes. Using the Hamiltonian (\ref{a1}) and taking into account the dissipation forces, one
readily obtains the following quantum Langevin
 equations:
 \small
 \begin{subequations}
 \begin{eqnarray} 
 &&\dfrac{dx}{dt}=\omega_m p,\\
 &&\dfrac{dp}{dt}=-\omega_m x-g_{_{1}}a^\dagger a -2g_{_{2}} a^\dagger a x - \gamma_m p+\xi(t),\label{a16}\\
 &&\dfrac{da}{dt}=-(\kappa+i\Delta) a -i g_{_{1}}a x-ig_{_{2}}  a x^2 +\varepsilon +\sqrt{2 \kappa}a_{in}, \\
 &&\dfrac{d(x^2)}{dt}=\omega_m (xp+px),\\
 &&\dfrac{d(p^2)}{dt}=-\left(\omega_m+2 g_{_{2}}a^\dagger a\right)(xp+px)-2 g_{_{1}}a^\dagger ap-2\gamma_m p^2\nonumber \\&&\hspace*{1cm}+2\gamma_m(1+2n_{th}),\\
 &&\dfrac{d(xp+px)}{dt}=-2 \left(\omega_m+2 g_{_{2}}a^\dagger a\right)x^2-2 g_{_{1}}a^\dagger ax+2 \omega_mp^2 \nonumber\\&&\hspace*{2cm}-\gamma_m(xp+px). \nonumber\\
 \end{eqnarray} \label{a2}\end{subequations}
 \normalsize where $n_{th}=[exp \left(\frac{\hbar \omega_m}{k_B T}\right)-1]^{-1}$ is the mean thermal phonon number.
 {\color {black}The mechanical mode, coupled to the thermal bath, is affected by a Brownian stochastic force described by $\xi(t)$ with zero-mean, having a damping rate $\gamma_m$ and the correlation function at temperature T as
 \small
\begin{equation}
 \langle \xi(t)\xi(t')\rangle= \frac{\gamma_m}{2\pi\omega_m}\int\omega e^{-i\omega(t-t')}\left[1+\coth\left(\frac{\hbar\omega}{2k_BT} \right)\right]d\omega, \label{a13}
\end{equation} \normalsize where  $k_B$ is the Boltzmann constant.}
 The input vacuum noise operator is represented as $a_{in}$ whose only non-zero correlation function is  
 \small
\begin{equation}
\langle \delta a_{in}(t)\delta a_{in}^\dagger(t')\rangle =\delta (t-t').\label{a14}
\end{equation}\normalsize
 We rewrite the Heisenberg operators as complex numbers, representing their respective steady state values
 with the inclusion of fluctuations around their steady state values.
 $i.e$ $\mathcal{O}(t)=\mathcal{O}_s+\lambda \delta \mathcal{O}(t)$. Expanding the set of equations Eqn.(\ref{a2}) in the manner described above leads us to a set of non-linear algebraic equations for steady state values, given by\small
 \begin{subequations}\begin{eqnarray}
 &&  x_s= \frac{-g_{_{1}}|a_s|^2}{\omega_m+2 g_{_{2}}|a_s|^2} ,
\label{a4}\\
 && (x^2)_s=\frac{g_{_{1}}^2|a_s|^4}{(\omega_m +2 g_{_{2}}|a_s|^2)^2}+\frac{\omega_m (1+2 n_{th})}{\omega_m +2 g_{_{2}}|a_s|^2},\\
 &&p_s= (xp+px)_s=0,\\
 &&(p^2)_s=1+2n_{th},\\
  &&  a_s  =\frac{\varepsilon}{ \kappa+i\left( \Delta+g_{_{1}}x_s +g_{_{2}}   (x^2)_s \right)}. \label{a3}
 \end{eqnarray}\label{a5}\end{subequations}\normalsize 
\vspace{-0.1cm}\subsection{Radiation Pressure and Quantum fluctuations}
 Since the fluctuations are assumed to be small when compared to the steady state values,
 we can neglect the non-linear terms in $\lambda$. This enables us to write the linearized
 Langevin equations for the fluctuations. \small
 \begin{subequations}
 \begin{eqnarray}
 \dfrac{d}{dt}\delta x &=&\omega_m \delta p,\\
 \dfrac{d}{dt}\delta p&=& -\omega_m \delta x -g_{_{1}} (a_s \delta a^\dagger+ a_s^* \delta a)-
 2 g_{_{2}}x_s(a_s^* \delta a+a_s \delta a^\dagger)\nonumber\\&&-2 g_{_{2}}|a_s|^2\delta x-\gamma_m \delta p+\xi(t),\\
 \dfrac{d}{dt}\delta a&=& -i\left(\Delta \delta a+g_{_{1}}\left( x_s \delta a+a_s \delta x\right)+ 
 g_{_{2}}x_s\left( 2a_s \delta x+x_s \delta a\right)\right)\nonumber\\&&-\kappa \delta a+\sqrt{2\kappa}\delta a_{in}.
  \end{eqnarray}\end{subequations}
   \normalsize
   By introducing 
 $\delta X=\frac{\delta a+\delta a^\dagger}{\sqrt{2}}$, $\delta P=\frac{\delta a-\delta a^\dagger}{\sqrt{2}i}$ and corresponding noises $\delta X_{in}$ and $\delta P_{in}$, we 
 rewrite the above equations in the compact form 
 \begin{equation}\dot{u}(t)=M u(t)+\nu(t),\label{a12}
\end{equation} with column vector of fluctuations in the system being $u^T= \left(\begin{matrix} \delta x,&\delta p,&\delta X,&\delta P \end{matrix}\right)$ and column vector of noise being 
$\nu^T=\left(\begin{matrix}0,&\xi(t),&\sqrt{2\kappa}\delta X_{in},& \sqrt{2\kappa}\delta P_{in}\end{matrix}\right)$. The matrix $M$ is given by
\small
\begin{eqnarray}M=\left(\begin{matrix} 0&\omega_m &0&0\\ -\tilde{\omega}_m& -\gamma_m& 
-\tilde{G} X_s&-\tilde{G} P_s\\ \tilde{G} P_s& 0&-\kappa  & \tilde{\Delta}\\ -\tilde{G} X_s & 0& -\tilde{\Delta}&-\kappa \end{matrix}\right) ,
\end{eqnarray}\normalsize
with  $I\equiv |a_s|^2$, $\tilde{\omega}_m \equiv \omega_m+2g_{_{2}} I$, $\tilde{\Delta}\equiv \Delta+g_{_{1}}x_s +g_{_{2}} x_s^2 $, $X_s=\frac{a_s+a_s^*}{\sqrt{2}}$, $P_s=\frac{a_s-a_s^*}{\sqrt{2}i}$ and
$\tilde{G} \equiv g_{_{1}}  +2 g_{_{2}}x_s$. 
{\color {black}The solutions of Eqn.(\ref{a12}) are stable only if all the eigenvalues of the matrix $M$, formed by the steady-state expectation values, have negative real parts. These stability conditions can be deduced by applying Routh-Hurwitz criterion \citep{routh} :}
\small
\begin{subequations}
\begin{eqnarray}
&&s_1 \equiv (\kappa^2+\tilde{\Delta}^2)+2 \kappa \gamma_m+\tilde{\omega}_m \omega_m >0,\\
&& s_2 \equiv (\kappa^2+\tilde{\Delta}^2)\gamma_m+2 \kappa \tilde{\omega}_m \omega_m>0,\\
&&s_3 \equiv (\kappa^2+\tilde{\Delta}^2)\tilde{\omega}_m \omega_m-\tilde{\Delta} \omega_m \tilde{G}^2 (X_s^2+P_s^2)>0,\\
&& (2 \kappa +\gamma_m) s_1 > s_2, \\
&& s_1 s_2 (2 \kappa +\gamma_m)> s_2^2+(2 \kappa +\gamma_m)^2 s_3.
\end{eqnarray}\label{a15}\end{subequations}\normalsize 
\par In experiments, fluctuations of the electric field are more convenient to measure in the frequency domain than in the time domain. Therefore by using the definition of Fourier transform, $\mathcal{F}(\omega)=\frac{1}{2\pi}\int_{-\infty}^\infty\mathcal{F}(t)e^{-i\omega t}dt$ and $[\mathcal{F}^\dagger(\omega)]^\dagger=\mathcal{F}(-\omega)$  in the Eqn.(\ref{a12}), the set of coupled differential equations form a simple system of linear equations in frequency.  Therefore after solving the matrix equation Eq.(\ref{a12}) in frequency domain, we get 
\small
\begin{eqnarray}
\hspace*{-.5cm}\delta a (\omega)=\frac{ A_a(\omega) \delta a_{in}(\omega)- A_{a^\dagger}(\omega) \delta a_{in}^\dagger (\omega) +A_{\xi}(\omega)\xi(\omega)}{ D(\omega)}, \label{b1}
\end{eqnarray} \normalsize 
with 
\small
\begin{subequations}
\begin{eqnarray}
A_a(\omega)&=&\sqrt{2\kappa}\left(-\omega_m\tilde{\omega}_m + \omega^2+i\omega\gamma_m\right)\left(\kappa-i(\omega+\tilde{\Delta})\right)\nonumber\\&&
+\sqrt{2\kappa} \omega_m\tilde{G}^2I,\\
A_{a^\dagger}(\omega)&=&i\sqrt{2\kappa}\omega_m(a_s^2\tilde{G}^2),\\
A_{\xi}(\omega)&=&i\omega_m\tilde{G}a_s(\kappa-i(\omega+\tilde{\Delta})),\\
D(\omega)&=& \left((\kappa -i \omega)^2+\tilde{\Delta}^2\right)\left( \omega^2+i \gamma_m \omega- \omega_m \tilde{\omega}_m)\right) \nonumber\\&&+2\tilde{G}^2I \tilde{\Delta}\omega_m .
\end{eqnarray} \label{b2}
\end{subequations}\normalsize 

\begin{figure*}[t]
\includegraphics[scale=0.33]{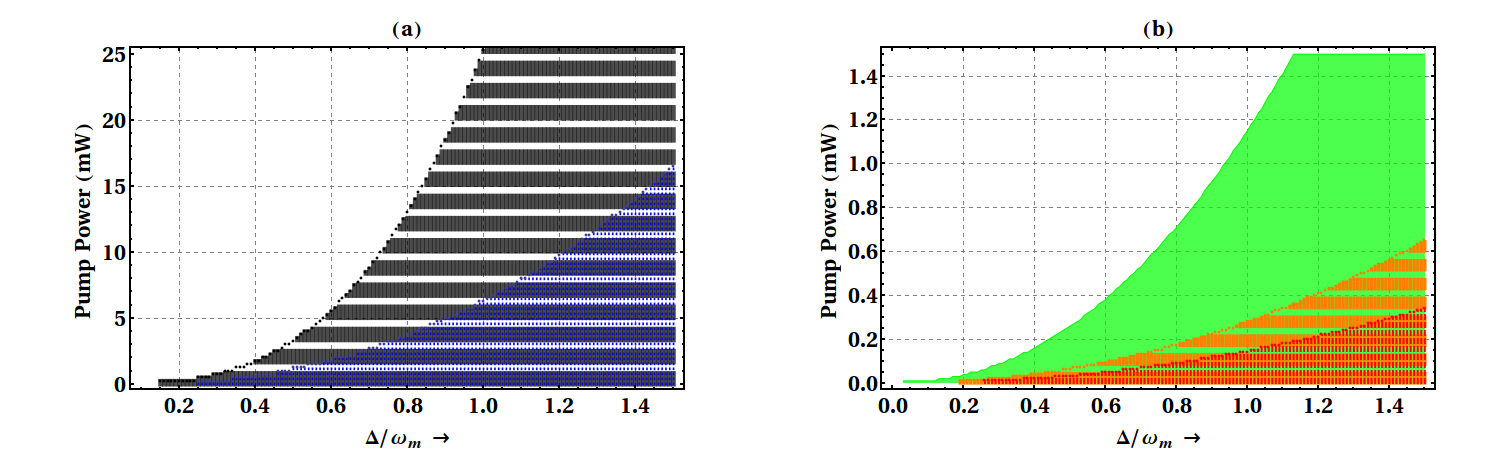}
                \caption{\label{fig2}{\color {black}(Color online)} Stability range of the system calculated as a function of normalised cavity detuning $\Delta/\omega_m$ and input power $\mathcal{P}$, for an optomechanical system with various QOC values ($g_{_{2}}/g_{_{1}}$). (a) The black striped region and blue dotted region correspond to $g_{_{2}}/g_{_{1}} =0$ and $-10^{-4}$ respectively. (b) The green shaded region, orange striped region and red dotted region correspond to  $g_{_{2}}/g_{_{1}}=-10^{-3}$, $-5\times 10^{-3}$ and $-10^{-2}$ respectively. Here $g_{_{1}}=1351.38$ Hz. }
\end{figure*}

The optical output field is related to the input field via the standard input-output relation $\delta a_{out}(\omega)=\sqrt{2\kappa}\delta a(\omega)-\delta a_{in}(\omega)$. This enables us to calculate the intensity spectrum of the cavity output-field given by 
\begin{equation}
S_{out}(\omega)=\frac{1}{2\pi}\int_{-\infty}^{\infty} \langle \delta a_{out}^\dagger(\omega')\delta a_{out}(\omega)\rangle e^{-i(\omega+\omega')t}d\omega'
\end{equation} To evaluate the spectrum, we need the correlations of noise operators in frequency domain defined as \small
\begin{subequations}
\begin{eqnarray}
&&\hspace*{-0.3cm}\langle \delta a_{in}(\omega)\delta a_{in}^\dagger(\omega')\rangle =2\pi\delta(\omega+\omega'),\\
&&\hspace*{-0.3cm}\langle \xi(\omega)\xi(\omega')\rangle = 2\pi \frac{\omega\gamma_m}{\omega_m}\left[\coth\left(\frac{\hbar \omega}{2k_BT}\right)+1\right]\delta(\omega+\omega').
\end{eqnarray}\label{b4}\end{subequations}\normalsize Using the above relations we get the  output-field spectrum as
\small \begin{eqnarray}
S_{out}(\omega)=&&\frac{2\kappa}{|D(\omega)|^2}\left\{ \frac{\omega\gamma_m}{\omega_m}\left(\coth\left[\frac{\hbar\omega}{2k_BT}\right]-1\right) \Big|A_{\xi}(\omega)\Big|^2  \right. \nonumber\\&&\left. \hspace*{4cm} +\Big|A_{a^\dagger}(\omega)\Big|^2\right\} \label{a18}
\end{eqnarray}\normalsize

 In Eqn. (\ref{a18}), the first term originates from the thermal noise of the mechanical oscillator, while the second term is from the cavity input vacuum noise. 

\subsection{Optical Quadrature squeezing spectrum}
 We use Eqn.(\ref{b1}) and (\ref{b2}) to analyse the squeezing of the optical quadrature as follows.
The squeezing spectrum of the cavity quadrature field is given by and calculated as \citep{shahidani}\small 
\begin{eqnarray}
S_\phi (\omega)&=&\dfrac{1}{4\pi}\int_{-\infty}^\infty d\omega' e^{-i(\omega+\omega')t}\langle \delta X_\phi^{out}(\omega)\delta X_\phi^{out}(\omega')+\nonumber\\&&\hspace*{1cm}\delta X_\phi^{out}(\omega')\delta X_\phi^{out}(\omega)\rangle ,\label{b3}
\end{eqnarray}\normalsize where $\delta X_\phi^{out}(\omega)= e^{-i\phi} \delta a_{out} (\omega) + e^{i\phi} \delta a_{out}^\dagger (\omega)$ is the Fourier transform of the output quadrature, with $\phi$ being its externally controllable quadrature phase angle that is experimentally measurable in a homodyne detection scheme \citep{loudon}. 

\par Using the Eqn.(\ref{b1}),(\ref{b2}) and (\ref{b4}) we can rewrite the squeezed spectrum of the cavity output-field as 
\small
\begin{eqnarray}
S_\phi(\omega)&=&C^{out}_{aa^\dagger}(\omega)+C^{out}_{a^\dagger a}(\omega)+e^{-2i\phi}C^{out}_{a a}(\omega)+e^{2i\phi}C^{out*}_{a a}(\omega).\nonumber\\
\end{eqnarray}\normalsize In the above equation $C_{aa}^{out*}(\omega)=C^{out}_{a^\dagger a^\dagger}(\omega)$ with the definition $ 2\pi \delta(\omega + \omega' )C_{\beta_1\beta_2}^{out} (\omega)\equiv \langle  \delta\beta_{1_{out}}(\omega) \delta\beta_{2_{out}} (\omega' )\rangle$ where $\beta_{1,2}$ can be $ a$ and $ a^\dagger$. 

We have to choose the values of the external parameters in order to achieve squeezing. Therefore we define optimum quadrature squeezing $S_{opt}(\omega)$ by choosing $\phi$ in such a away that  $dS_\phi (\omega)/d\phi = 0$, yielding us
\small
\begin{equation}
e^{2i\phi_{opt}} = \pm\dfrac{C^{out}_{aa}(\omega)}{|C^{out}_{aa}(\omega)|} .
\end{equation}\normalsize
We need to choose the solution with a negative sign as it minimises the spectrum function. Therefore we have,
\small \begin{eqnarray}
S_{opt}(\omega)&=&C^{out}_{a a^\dagger}(\omega)+C^{out}_{a^\dagger a}(\omega)-2|C^{out}_{a a}(\omega)|, \label{sopt}
\end{eqnarray}\normalsize
where \footnotesize
\begin{subequations}
\begin{eqnarray}
&&\begin{split}
C^{out}_{aa}(\omega)&=\frac{2}{|D(\omega)|^2} \bigg\{\kappa\left(\dfrac{2\omega\gamma_m}{\omega_m}A_\xi(\omega)A_\xi(-\omega)\coth\left[\dfrac{\hbar\omega}{2k_BT}\right]\right.\\ &+A_a(-\omega)A_{a^\dagger}(\omega)  +A_a(\omega)A_{a^\dagger}(-\omega)\bigg)\\&-\sqrt{\dfrac{\kappa}{2}}\big( D(\omega)A_{a^\dagger}(-\omega)+D(-\omega)A_{a^\dagger}(\omega)\big)\bigg\},
\end{split}\nonumber\\\end{eqnarray}
\begin{eqnarray}
&&\begin{split}
C^{out}_{a^\dagger a}(\omega)&=\frac{2}{|D(\omega)|^2} \bigg\{\dfrac{\kappa\omega\gamma_m}{\omega_m}\left(|A_\xi(\omega)|^2\left(1+\coth\left[\dfrac{\hbar\omega}{2k_BT}\right] \right)\right. \\&+\left.|A_\xi(-\omega)|^2\left(-1+\coth\left[\dfrac{\hbar\omega}{2k_BT}\right] \right)\right) +\kappa\Big(|A_{a^\dagger}(\omega)|^2 \\& +|A_{a^\dagger}(-\omega)|^2\Big)\bigg\}.
\end{split}\nonumber\\\\
&&\begin{split}
C^{out}_{aa^\dagger}(\omega)&=\frac{2}{|D(\omega)|^2} 
\bigg\{\dfrac{\kappa\omega\gamma_m}{\omega_m}\left(|A_\xi(-\omega)|^2\left(1+\coth\left[\dfrac{\hbar\omega}{2k_BT}\right] \right)\right. \\&+\left.|A_\xi(\omega)|^2\left(-1+\coth\left[\dfrac{\hbar\omega}{2k_BT}\right] \right)\right) +\kappa \Big(|A_{a}(\omega)|^2 \\& +|A_{a}(-\omega)|^2\Big)-\sqrt{\dfrac{\kappa}{2}}\Big( D(-\omega)\big(A^*_{a^\dagger}(-\omega)+A_a(\omega)\big)\\&+D(\omega)\big(A^*_{a^\dagger}(\omega)+A_a(-\omega)\big)\Big)\bigg\}+2.
\end{split}\nonumber\\
\end{eqnarray}\label{cout}
\end{subequations}
\normalsize
The squeezing occurs for the condition, $S_{opt} (\omega) < 1$ and the quadrature operator commutation relation $[\delta X^{out}_\phi(\omega),\delta X^{out}_{\phi+\pi/2}(\omega)] = -2i$ has to be satisfied. The $S_{opt}(\omega) =1$ represents the spectrum of fluctuation for a coherent or vacuum field and $S_{opt} (\omega) = 0$ corresponds to a perfect squeezing.
To understand the parameter dependence on the amount of squeezing, we consider the limits $ \omega_m>\kappa\gg\gamma_m$ at $T=0$ K and derive a simple expression for $S_{opt}(\omega)$. Using this expression one can estimate the amount of squeezing. The results are available at the end of this article, in the Appendix section.  
\begin{figure}[t]
 \includegraphics[scale=.29]{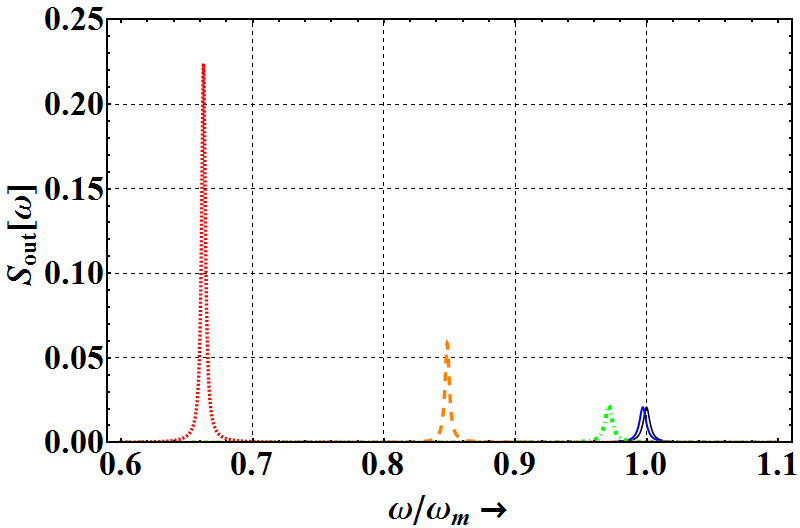}
                \caption{\label{fig3}{\color {black}(Color online)}The spectrum of the cavity output-field plotted as a function of normalised frequency ($\omega/\omega_m$) for an optomechnical system with different QOCs. The system parameters are $\Delta = \omega_m$, $T=1$ mK, and $\mathcal{P}=100$ \textmu W.  The curves black (thick solid), blue (solid), green (dot-dashed), orange (dashed) and red (dotted) correspond to QOC values as $g_{_{2}}/g_{_{1}} =0$, $-10^{-4}$, $-10^{-3}$, $-5\times 10^{-3}$ and $-10^{-2}$ respectively with $g_{_{1}}=1351.38$ Hz.}
\end{figure}   

\section{Numerical results and discussion}
In our calculations we choose the  parameters similar to those used in \citep{vitali2007optomechanical}: cavity length $L=1$ mm and driven by a laser of wavelength 810 nm . The mechanical oscillator  has a frequency $ \omega_m=2\pi\times 10$ MHz, damping rate $\gamma_m=2\pi\times 100$ Hz and mass $m=5$ ng. The linear optomechanical coupling rate, $g_{_{1}}$ is estimated as $\dfrac{\partial \omega_c}{\partial x} x_{zpf} =1351.38$ Hz. In order to work in the resolved side-band limit we choose cavity damping rate as $\kappa=2\pi \times 1$ MHz. To analyse the effects of QOC on this system, we fix LOC and vary QOC. In all our numerical results we scale QOC values with LOC value shown as $g_{_{2}}/g_{_{1}}$.

 Considering  LOC and QOC together in the system provides  highly non-linear interaction between cavity light field and mechanical motion.
 The system is said to be stable when the two forces, radiation pressure force and the mechanical restoring force, balance each other. The restoring force depends on the spring constant of the mechanical oscillator. From Eqn.(\ref{a16}) the effective spring constant of the mechanical oscillator can be written as  $\frac{\hbar}{x_{zpf}^2}\left( \omega_m+2g_{_{2}}I\right)=m\omega_m\tilde{\omega}_m$. The presence of QOC and its value either positive or negative will stiffens or softens the spring respectively. Here we have a negative QOC which softens the spring, decreasing the maximum restoring force. Hence to balance this modified restoring force the radiation pressure force that is given by $F_{rad}= (\hbar \omega_c/L)\langle a^\dagger a\rangle\propto\frac{\mathcal{P}}{\tilde{\Delta}^2+\kappa^2}$, should also decrease accordingly. Therefore the stability range when mapped as a function of detuning and power gets smaller as QOC  increases.
 \begin{figure}[t]
 \includegraphics[scale=.39]{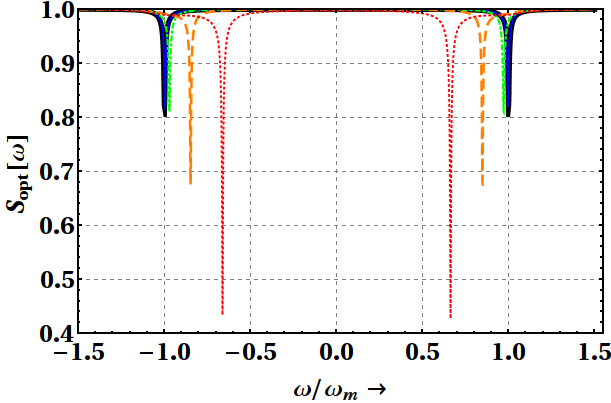}
                \caption{\label{fig4}{\color {black}(Color online)}
Squeezing spectrum for cavity quadrature field plotted as a function of normalised frequency ($\omega/\omega_m$) for an optomechnical system with different QOCs. The system parameters are same as in Fig. \ref{fig3}.}
\end{figure}
\begin{figure*}[t]
 \includegraphics[scale=.4]{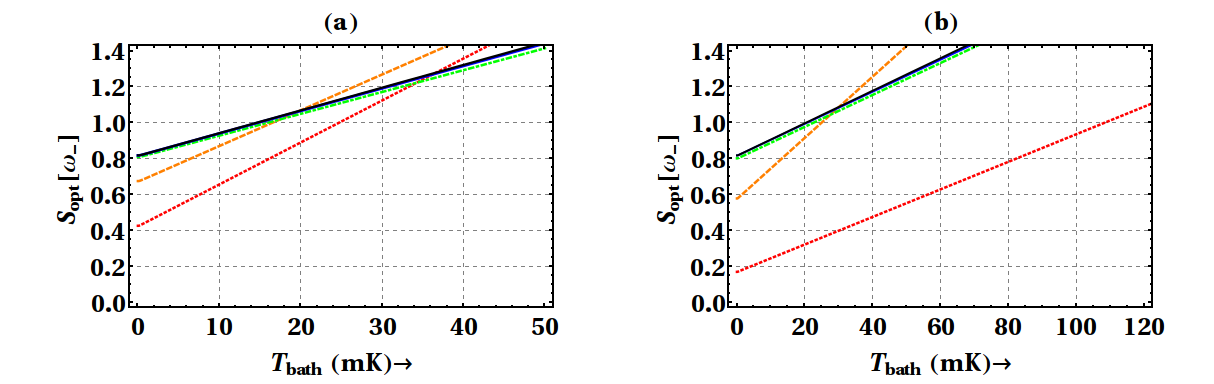}
                \caption{\label{fig5}(a), (b){\color {black}(Color online)} shows squeezing spectrum for cavity quadrature field evaluated at $\omega_-$ and varied as a function of bath temperature for different QOCs with powers 100 \textmu W and 140 \textmu W respectively. The colour code and all other system parameters are same as in Fig. \ref{fig3}. }
\end{figure*}

 We show this by calculating the intensity circulating inside the cavity, $I=|a_s|^2$ at low thermal phonon numbers using the approximation $ (x^2)_s\approx \left( x_s\right)^2$ in Eqn.(\ref{a5}). Then using the Routh-Hurwitz criteria Eqn.(\ref{a15}), the stability of the system is deduced. Fig. \ref{fig2} displays stability range of the system over input power ($\mathcal{P}$) and normalised cavity detuning ($\Delta/\omega_m$) for different QOCs. In Fig. \ref{fig2} the black striped region correspond to system with no QOC and other coloured regions belong to system with various QOCs. It is clear that in the former case the stability range is far bigger than the latter with power decreasing from mW to \textmu W respectively. 

 The presence of QOC also modifies the mechanical oscillator's frequency, $\omega_m$ and single photon optomechanical coupling, $g_{_{1}}$ to their intensity dependent counterparts $\sqrt{\omega_m\tilde{\omega}_m}$ and $\tilde{G}$ respectively. These modifications affect the spectrum of the output-field that is reflected from the oscillating mirror. While a change in the optomechanical coupling  enhances the amplitude of the fluctuations , the modified oscillator's frequency results in a shift in the spectral peaks.    Using Eqns. (\ref{b2}) and (\ref{a18}), the spectrum of cavity output-field is plotted as a function of normalized frequency ($\omega/\omega_m$). The black curve corresponds to LOC alone where as blue, green , orange and red correspond to  various QOCs as shown in Fig. \ref{fig3}. With QOC an increase of ten folds (red curve in Fig. \ref{fig3}) the amplitude of output-field spectrum can be observed as compared to LOC (black curve in Fig. \ref{fig3}). Further, the spectral features of the cavity output-field $S_{out}(\omega)$ can be analysed by studying $D(\omega)$. The real and imaginary parts give the position and width of the spectral peaks respectively. The peaks occur at the frequencies centred at $\omega_+$ around $\pm\sqrt{\tilde{\Delta}^2+\kappa^2}$ and $\omega_-$ around $\pm\sqrt{\omega_m\tilde{\omega}_m}$.\footnote{Note that the expression for frequencies $\omega_\pm$ are just an approximation. An explicit expression is given in the Appendix.} The shift in the spectral peaks is clearly manifested for higher values of QOC (orange and red curve) as shown in Fig. \ref{fig3}. Since it is evident that the shifted and enhanced spectral features of the cavity output-field is observed at these frequencies, one expects the same to follow for the squeezing spectrum too. 

 It has been shown in \citep{mancini,fabre1994quantum} that an optomechanical system mimics as a kerr medium that can generate non-classical light. Since in our scheme the single photon optomechanical coupling is effectively replaced by the bigger intensity dependent coupling $\tilde{G}$, we reckon that the system acts as an enhanced kerr medium. To achieve ponderomotive squeezing it is necessary that the radiation pressure force noise has to dominate over the effects of thermal noise. From Eqn. (\ref{b2})  it is clear that the contribution of thermal noise term $A_\xi (\omega) \propto \tilde{G}a_s$ and  contribution from the input vacuum fluctuations $A_{a^\dagger} (\omega)$ and $A_a(\omega)$ are proportional to  $\tilde{G}^2a_s^2 $ and $\tilde{G}^2I$ respectively, facilitates  an enhancement in the amount of  squeezing of the output field around the modified frequency of mechanical oscillator $\sqrt{\omega_m\tilde{\omega}_m}$.

To show this, using Eqn.(\ref{sopt}) and (\ref{cout}) squeezing spectrum for cavity quadrature field as a function of normalized frequency ($\omega/\omega_m$) at 1 mK as the bath's temperature and 100 \textmu W pump power is calculated and plotted in Fig. \ref{fig4}. Higher squeezing is obtained in presence of QOC as compared with LOC. 
 In all our numerical calculations, a considerable amount of squeezing is obtained at $\omega_-$. We observe that the effect of QOC is remarkable that bearing a value as low as 1/100 times (red curve) of LOC could aid in enhancing squeezing thrice ($\sim$  60\%) as compared to its LOC (black curve) counterpart ($\sim$ 20\%).

It is also important to study the effect of bath's temperature on the system. The presence of thermal phonons limit the achievable level of squeezing in the system. This is evident in Fig. \ref{fig5}(a), (b) where we plotted squeezing spectrum evaluated at $\omega_-$ for $\mathcal{P}=100$ \textmu W  and $\mathcal{P}=140$ \textmu W respectively. In Fig. \ref{fig5}(a), as we increase temperature the achievable level of squeezing degrade for every curve. This results in the temperature $T_c$, only below which for a fixed system parameters and laser input power, squeezing occurs \textit{i.e.}, $S_{opt}<1$.  It is important to notice that, the system with QOC (red,  orange, green and blue curves) have a $T_c$ higher than that of the system with LOC interaction alone (black curve).  
Figure. \ref{fig5}(b) shows a similar plot but at higher power. Increasing input power enhances the maximum level of  squeezing attained in the presence of QOC more when compared to a LOC system, at any bath temperature below $T_c$. For example, one can see that the squeezing has enhanced by three folds (20\% to 60\%) using the value of $g_{_{2}}/g_{_{1}}=-10^{-2} $ at 1 mK in Fig. \ref{fig5}(a) and by four folds (20\% to 80\%) in Fig. \ref{fig5}(b).

It is also interesting to find that the $T_c$ can be increased considerably by increasing input power a little, when QOC is included. This results in squeezing of the optical quadrature at temperatures where LOC systems cannot exhibit squeezing . This is quiet evident in the Fig. \ref{fig5}(b)in the range of 20 mK to 110 mK. Unlike the system with LOC interaction alone (black curve),  we find that at 20 mK  nearly  65\% squeezing can be obtained with $g_{_{2}}/g_{_{1}}=-10^{-2}$. This would be of a greater importance for experimentalists as one can achieve temperatures of 20 mK using dilution refrigerators. Though higher input powers with a higher value of QOC increases $T_c$ and thereby the level of squeezing, it is equally important to know how  QOC can affect the necessary input power required to squeeze the optical quadrature. 

\begin{figure}[t]
                       \includegraphics[scale=.36]{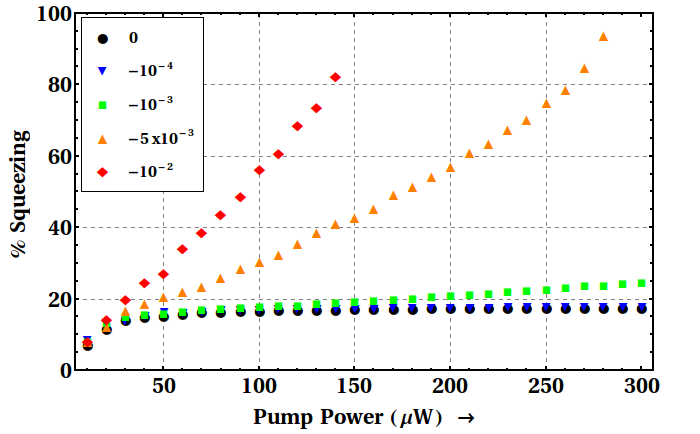}
               \caption{\label{fig6}{\color {black}(Color online)}
 Percentage of squeezing evaluated at $\omega_-$ and varied as a function of pump power $\mathcal{P}$ for different QOC values ($g_{_{2}}/g_{_{1}}$) as shown in the legend. The system parameters are same as in Fig. \ref{fig3}.
}
\end{figure}

The Fig. \ref{fig6} shows percentage of quadrature squeezing evaluated at $\omega_-$ and plotted as a function of $\mathcal{P}$ for various QOCs at 1 mK. For low QOC value (blue) the achievable squeezing percentage looks the same as that of the LOC system (black curve). The prominence of QOC  on squeezing increases immensely as we increase its value. Even at the value where QOC (red curve) is 100 times lower than LOC, we see a significant rise of percentage of squeezing as function of power. This suggests that the inclusion of QOC can aid in achieving higher levels of squeezing at lower powers.
 
 The crucial factor that governs the maximum level of squeezing depends on the value of QOC and also on system's stability range. Both these have to be optimized in such a way that squeezing percentage becomes maximum. From the Fig. \ref{fig6} it is clear that though the curve with higher QOC (red curve  corresponding to $g_{_{2}}/g_{_{1}}=- 10^{-2}$) shows a sharper rise of squeezing percentage, the orange curve corresponding to $g_{_{2}}/g_{_{1}}=-5\times 10^{-3}$ gives the maximum level of squeezing  $>$ 90\%. This is due to the fact that beyond $\mathcal{P}=140$ \textmu W there are no stable states in the system for $g_{_{2}}/g_{_{1}}=- 10^{-2}$ (red curve) where as the former has its stability range till 280 \textmu W as shown in Fig. (\ref{fig2}).  Therefore an optimal QOC has to be chosen which is large enough to generate optical quadrature squeezing at lower pump powers and can provide a larger stability range for the system. These two features together yield us a maximum level of squeezing possible in the system. 

\section{Conclusions}
In summary we have analysed  and compared a cavity-optomechanical system containing LOC and QOC together to a system with LOC alone. In doing this, we successfully showed that the inclusion of QOC to a conventional LOC optomechanical system shrinks the stability range and can aid in higher squeezing at lower powers and higher bath temperatures. We also found that at a particular bath temperature, having a higher value of QOC does not necessarily lead to higher percentage of achievable squeezing. There has to be a trade-off  between the QOC and stability range such that one can maximise the squeezing. Unlike hybrid-optomechanical systems \citep{shahidani,eyob,electromechanical}, the system under study is advantageous as one can preserve all the capabilities of controlling the level of squeezing without introducing any extra degrees of freedom. Our proposed scheme can be implemented in already existing systems like \citep{purdy,xuereb,j-d-thompson}. The present study demonstrates that, systems with both LOC and QOC interactions can be a good platform for achieving higher squeezing in accessible parameter space and hence can be used in realizing better quantum mechanical devices.
\section*{Acknowledgement} We thank Dr. Sachin Kasture, Dr. Sourav Dutta and  Dr. Andal Narayanan for going through the manuscript and giving us their valuable suggestions. 

\appendix*
\section{}
\subsection{Explicit expression for $\omega_\pm$}
First we present the explicit expression for the frequencies at which spectral peaks (squeezing) occurs. This can be calculated by finding the roots of the real part of  $D(\omega)$ as written in Eqn. (\ref{b2}).  These are
\small \begin{eqnarray}
&&\omega^2_\pm=\frac{1}{2}\left(\kappa^2+\tilde{\Delta}^2+2\kappa\gamma_m+\omega_a\right)\nonumber\\&&\pm\frac{1}{2}\sqrt{\left(\kappa^2+\tilde{\Delta}^2+2\kappa\gamma_m+\omega_a\right)^2+4\left(2\omega_mI\tilde{G}^2\tilde{\Delta}-\left(\kappa^2+\tilde{\Delta}^2\right)\omega_a\right)}\nonumber\\\label{A1}
\end{eqnarray}\normalsize 
with $\omega_a\equiv\sqrt{\omega_m\tilde{\omega}_m}$.
\subsection{Parameter dependence of $S_{opt}(\omega)$}
 Here we present our analysis and result to approximate the output squeezed spectrum in the limit $$
T=0,\quad\gamma_m =0,\quad \tilde{\Delta}\approx \omega_m > \kappa, $$
that allows us to write $ \quad \coth\left[\dfrac{\hbar \omega}{2 k_BT}\right]=1$ and $X_s^2+P_s^2 \approx P_s^2=2I$ with $X_s \approx 0$. 

We redefine Eqn. (\ref{b2}) in the above limit as they form the expression $S_{opt}(\omega)$. We eliminate $\gamma_m$ in all the expressions and replace $\kappa \pm \tilde{\Delta}$ and $\kappa \pm \omega$ with $\tilde{\Delta}$ and $\omega$ respectively in $A_{\xi}(\omega),A_a(\omega)$ and $A_{a^\dagger} (\omega)$. We retain the multiplicative linear term in $\kappa$ i.e. $2i\kappa\omega$ in $D(\omega)$  and ignore the higher order terms of $\kappa$. Thereafter we get a simpler expression for in system parameters.
\begin{widetext}
\small
\begin{eqnarray}
S_{opt}(\omega)=1+\dfrac{8\kappa^2 \omega_m\tilde{G}^2I\left(\omega_m\tilde{G}^2I+\tilde{\Delta}\left(\omega^2-\omega_a^2\right)\right)+4\kappa^2\left(\omega^2-\omega_a^2\right)^2\left(\tilde{\Delta}^2-\omega^2\right)}{|D(\omega)|^2}\nonumber\\-\dfrac{4\kappa \omega_m\tilde{G}^2I\Big|2\tilde{G}^2I\omega_m \left(\kappa-i\tilde{\Delta}\right)+i\left(\omega^2-\omega_a^2\right)\left(\omega^2-\tilde{\Delta}^2-2i\kappa\tilde{\Delta}\right)\Big|}{|D(\omega)|^2} \label{a17}
\end{eqnarray} \normalsize\end{widetext}
 where, \\$D(\omega)=\left(\omega^2-\omega_a^2\right)\left(\tilde{\Delta}^2-\left(\omega^2+2i\kappa\omega\right)\right)+2\omega_m\tilde{\Delta}\tilde{G}^2I$ \normalsize 
 
 The above expression can be used to estimate the amount of squeezing by using the appropriate solution, \textit{i.e.} $\omega_-$ from Eqn. (\ref{A1}) for $\omega$.

\bibliography{ref}
\end{document}